\RequirePackage{lineno} 
\documentclass[%
letterpaper,
reprint,
superscriptaddress,
showpacs,
amsfonts,amsmath,amssymb,
aps,
prb,
floatfix
]{revtex4-1}
\usepackage{graphicx}
\usepackage{dcolumn}
\usepackage{bm}

\begin{document}
\title{Defect physics and electronic properties of Cu$_3$PSe$_4$ from first principles}

\author{D.~H.~Foster}
\affiliation{Department of Physics, Oregon State University, Corvallis, Oregon 97331, USA}

\author{F.~L.~Barras}
\affiliation{Department of Physics, Oregon State University, Corvallis, Oregon 97331, USA}

\author{J.~M.~Vielma}
\affiliation{Department of Physics, Oregon State University, Corvallis, Oregon 97331, USA}

\author{G.~Schneider}
\email{Guenter.Schneider@physics.oregonstate.edu}
\affiliation{Department of Physics, Oregon State University, Corvallis, Oregon 97331, USA}

\date{\today}

\begin{abstract}
The $p$-type semiconductor Cu$_3$PSe$_4$~has recently been established to have a direct bandgap of 1.4 eV and an optical absorption spectrum similar to GaAs [Applied Physics Letters, 99, 181903 (2011)], suggesting a possible application as a solar photovoltaic absorber.
Here we calculate the thermodynamic stability, defect energies and concentrations, and several material properties of Cu$_3$PSe$_4$~using a wholly GGA+$U$ method (the generalized gradient approximation of density functional theory with a Hubbard $U$ term included for the Cu-$d$ orbitals).
We find that two low energy acceptor defects, the copper vacancy V$_{\text{Cu}}$ and the phosphorus-on-selenium antisite P$_{\text{Se}}$, establish the $p$-type behavior and likely prevent any $n$-type doping near thermal equilibrium.
The GGA+$U$ defect calculation method is shown to yield more accurate results than the more standard method of applying post-calculation GGA+$U$-based bandgap corrections to strictly GGA defect calculations.
\end{abstract}


\maketitle

\section{Introduction}
The growing family of multinary copper chalcogenides has been of great interest for solar photovoltaic applications.
In addition to the commonly used solar absorber CuIn$_{1-x}$Ga$_x$Se$_2$ (CIGS), materials that have raised interest include Cu$_2$ZnSnS$_4$, Cu$_7$TlS$_4$\cite{yu12ide}, CuClSe$_2$\cite{wang09pre}, CuBiS$_2$\cite{dufton12str}, CuSbS$_2$\cite{dufton12str}, and Cu$_3$BiS$_3$\cite{colombara12for}.
Recently the $p$-type semiconductor Cu$_3$PSe$_4$~has been established\cite{foster11ele} to have a direct bandgap of $E_{g} = 1.4$ eV, with a calculated absorption $\alpha > 5\times10^4$ cm$^{-1}$ for wavelengths less than $630$ nm.
This bandgap lies in the optimal range for photovoltaic power output and warrants further investigation of the material.

In addition to optical absorption, essential considerations for photovoltaic applications include ease of synthesis, conductivity, amenability to doping, and trap-assisted charge recombination.
These quantities are largely controlled by the thermodynamic stability of the material with respect to competing phases and point defects.
Materials which allow bipolar doping (both $n$-type and $p$-type behavior are achievable through doping) are of special interest because $p$-$n$ homojunction capability may reduce the number of heterojunctions needed in a solar cell design.
Bipolar doping occurs under typical synthesis techniques only when all intrinsic defects have charge transition energies and formation energies large enough so that extrinsic (dopant) charged defect states are energetically favorable for a sizable range of Fermi energies, extending well above and below the center of the bandgap.
Computational defect analysis using relatively inexpensive methods can often determine with good confidence whether bipolar doping is possible for synthesis methods near thermal equilibrium.

Here we perform a point defect analysis of Cu$_3$PSe$_4$~combining the +$U$ Hubbard term for total energy calculations with the correction methods described recently by Lany and Zunger\cite{lany08ass, lany09acc}.
Several potential substitutional donor defects are also considered.
Furthermore we examine bulk properties including the partial density of states (DOS), the dielectric tensor, and the highly asymmetric effective mass tensor.
We compare our results to recent experiments\cite{kokenyesi12tbd} and to a more standard procedure using the generalized gradient approximation (GGA) for defect supercell calculations followed by a post-calculation valence band correction.
We also compare our methods with the alternative electrostatic image correction procedure described by Freysoldt \textit{et al}.\cite{freysoldt10ele,*freysoldt09ful}.

\section{Methods}
\subsection{Computation}

Defect formation energies are most often calculated using density functional theory (DFT) within the local density approximation (LDA) or within the GGA.
However, recent statistical studies\cite{lany08sem, stevanovic12cor} on the accuracy of heat of formation calculations indicate that using GGA with an additional Hubbard $U$ term for the occupation of transition metal $d$ orbitals, the so-called GGA+$U$ method, will be more accurate than using standard GGA or LDA.
Furthermore, the defect study by Scanlon \emph{et al}.\cite{scanlon09mod} has compared the GGA and GGA+$U$ methods for $V_{\text{Cu}}$ and $I_{\text{O}}$ defects in Cu$_2$O, and found that in comparison to the GGA with a valence band correction, the wholly GGA+$U$ method reproduced more (although not all) of the experimental features sought.
In the GGA+$U$ method, the $U$ value is held constant for each type of transition metal atom throughout the analysis, including calculations of the energies of the transition metal elements themselves.

The heat of formation studies\cite{lany08sem, stevanovic12cor} also suggest that one should add a statistically determined correction value to the total energy of each pure element before calculating the heat of formation $\Delta H$ of a compound.
To obtain the most accurate heat of formation energies for both compounds and defects, we use GGA+$U$\cite{dudarev98ele} and apply the elemental energy corrections suggested by Lany\cite{lany08sem} for P in all phosphides\cite{Note1} and for Ca in all Ca compounds.
The other elements we consider, Cu, Se, Zn, Cd, and Cl, either have statistically insignificant corrections or, in the case of Cl, are not considered in Ref.~\onlinecite{lany08sem}.

Our calculations use the projector augmented wave (PAW) method\cite{bloechl94pro,kresse99fro} as implemented in the plane wave code \texttt{VASP}\cite{kresse96eff} with the Perdew-Burke-Ernzerhof\cite{perdew96gen} (PBE) parameterization of the GGA exchange-correlation functional.
We use an effective $U$ value of 6 eV for the \hbox{Cu-$d$}, \hbox{Zn-$d$}, and \hbox{Cd-$d$} orbitals.
This value of $U$ for \hbox{Cu-$d$} has been chosen in previous work (c.f.~Ref.~\onlinecite{persson05ntd}) to yield agreement with the experimental band structure below the valence band maximum (VBM)\cite{Note2}, thus eliminating or significantly reducing the need for post-calculation corrections to the VBM of Cu$_3$PSe$_4$.\cite{Note3}
Calculations use a plane wave cutoff energy of \hbox{310 eV}
and a set of comparison calculations using cutoff energy \hbox{400 eV} resulted in very small corrections of order \hbox{0.01 eV}.
The density functional perturbation theory calculations we report below were calculated with a \hbox{400 eV} cutoff.
In the image charge corrections and the hydrogenic binding energy estimations, we have used the value $\epsilon_0=14.1$, which was calculated with the \hbox{310 eV} cutoff.
 All calculations include ionic relaxation, while lattice parameters are relaxed for all pure compounds and elements, including the defect free host.
Lattice parameters are determined by performing shape relaxations for a sequence of cell volumes, and interpolating the volume of minimum energy using the Murnaghan equation of state.
Perturbation of ions is used to destroy symmetry within the supercell calculations.
We primarily use $2 \times 2 \times 2$ ($2^3$) supercells ($\sim\! 128$ atoms) with a \hbox{$\Gamma$-centered} $2^3$ \hbox{$k$-point} grid.

The analysis and correction methods used here are chosen in an attempt to maximize accuracy without entailing a much more costly analysis using more accurate electronic structure methods, such as hybrid functionals.
For fixed, experimental lattice parameters, we have compared bulk Cu$_3$PSe$_4$ calculations for GGA, GGA+$U$, and the Heyd-Scuseria-Ernzerhof\cite{heyd_hybrid_2003,*heyd_erratum:_2006} (HSE) hybrid functional.
We find that for a number of properties, including the P-Se and Cu-Se bond lengths and the lowest conduction band charge distribution, the GGA+$U$ results are significantly closer to the HSE results than the GGA results.
The HSE functional itself yields unexpectedly accurate results for the bandgap (error $\approx 0.02$ eV) and bond lengths (error $\approx 0.01$ \AA)\cite{foster11ele}.
The similarity of GGA+$U$ and HSE bulk calculations thus raises our expectations for GGA+$U$ performance, particularly for shallow acceptor defects which should avoid bandgap related uncertainties when $\Delta E_V = 0$.

\subsection{Defect Heat of Formation}

The defect formation energies are performed using the formula
\begin{align}
\Delta H_{D,q}(&E_F,\{\Delta \mu_{\alpha} \}) =  E_{D,q} - E_H + (E_V + E_F)q \notag \\
&+ \sum_{\alpha}  (\mu_{\alpha}^0 + \Delta \mu_{\alpha})n_{\alpha} + \Delta E_{\text{corr}} .
\end{align}

The notation here follows Ref.~\onlinecite{lany08ass}: $D$ denotes the defect type, $q$ is the charge of the defect charge state, $E_F$ is the Fermi energy level, $E_V$ is the host VBM, $E_H$ is the calculated total energy of the host supercell, and $E_{D,q}$ is the calculated total energy of the defect supercell. 
$\mu_{\alpha} \equiv  \mu_{\alpha}^0 + \Delta \mu_{\alpha}$ is the chemical potential for atom type $\alpha$ in the synthesis environment with $\mu_{\alpha}^0$ being the calculated pure element energy (possibly with statistical corrections\cite{lany08sem,stevanovic12cor}) and $\Delta \mu_{\alpha} \leq 0$ being determined by synthesis conditions.
$n_{\alpha}$ is the number of atoms added to the environment in creation of the defect $D$.

The energy correction term $\Delta E_{\text{corr}}$ is expanded as 
\begin{align}
\Delta E_{\text{corr}} =& \Delta E_{\text{BF}} + \Delta E_{\text{PA}} + \Delta E_{\text{MP}} \notag \\
& \quad + q \Delta E_V - z_h \Delta E_V + z_e \Delta E_C . \label{e:Ecorr}
\end{align}
The meaning of the last three (bandgap correction) terms follows the description in Ref.~\onlinecite{persson05ntd} except that $\Delta E_V$ is defined here to be positive for a
\hbox{gap-narrowing} correction.
For the GGA+$U$ defect calculations, we assume $\Delta E_V = 0$, while the correction for the GGA calculations is described in Appendix \ref{as:GGAVBMcorr}.
$\Delta E_C$ is the correction to the conduction band minimum (CBM), determined from $\Delta E_V$ and the experimental and calculated band gaps of the host.
For reasons discussed below, we only apply the shallow donor correction $z_e \Delta E_C$ for the extrinsic shallow donor defects such as Zn$_{\text{Cu}}^0$.
For all other defects we take the ``band edge only'' approach to the conduction band correction, in which we do not change the transition energies as $E_C$ is moved.
Here $z_e$ is the number of electrons locally bound in a shallow donor state and $z_h$ is the number of holes locally bound in a shallow acceptor state. 

The first three terms of Eq.~(\ref{e:Ecorr}) are the band filling correction, the potential alignment correction, and a modified Makov-Payne electrostatic image correction, respectively.
These terms collectively are the finite size correction terms, and they follow Refs.~\onlinecite{lany08ass, lany09acc}.

The band filling correction for the acceptor defects is given by
\begin{align}
\Delta E_{\text{BF}}(D,q) =&  - \sum_{n,\bm{k}} w_{\bm{k}} (2 - \eta_{n,\bm{k}}) (\tilde{e}_V - e_{n,\bm{k}} ) \notag \\
& \quad \times \Theta (\tilde{e}_V - e_{n,\bm{k}} ) , \label{e:bf}
\end{align}
where $\Theta(x)$ is the Heaviside step function, $w_{\bm{k}}$ is the \hbox{$k$-point} weight, $\eta_{n,\bm{k}}$ is the occupancy of the two-electron state ($n,\bm{k}$), $e_{n,\bm{k}}$ denotes the state eigenvalue, and $\tilde{e}_V$ is the host VBM adjusted by the potential offset:
\begin{align}
\tilde{e}_V = E_{V,H} + (V_{D,q}^r - V_{H}^r) .
\end{align}
The potential references $V^r$ are calculated by averaging the atomic 
sphere-averaged core potentials excluding the defect site, and in some cases nearest neighbor sites, as described in Ref.~\onlinecite{lany09acc}.

We have not included dispersion corrections to isolated, half occupied \textit{deep} defect states, since these corrections are found to be small in light of the much larger uncertainties of deep state transition energies.

The potential alignment correction is given by
\begin{align}
\Delta E_{\text{PA}} = q (V_{D,q}^r - V_{H}^r).
\end{align}

The modified Makov-Payne correction is $2/3$ multiplied by the monopole ($1/L$) term,
\begin{align}
\Delta E_{\text{MP}} = \frac{2}{3} \frac{q^2 \alpha_M}{2 \epsilon_0 L} ,
\end{align}
as derived in Ref.~\onlinecite{lany08ass}.
For the $2^3$ Cu$_3$PSe$_4$~supercell, $|q|=1$, and $\epsilon_0=14.1$, we find $\Delta E_{\text{MP}} = 0.069$ eV.

\begin{figure*}[!t]
\includegraphics{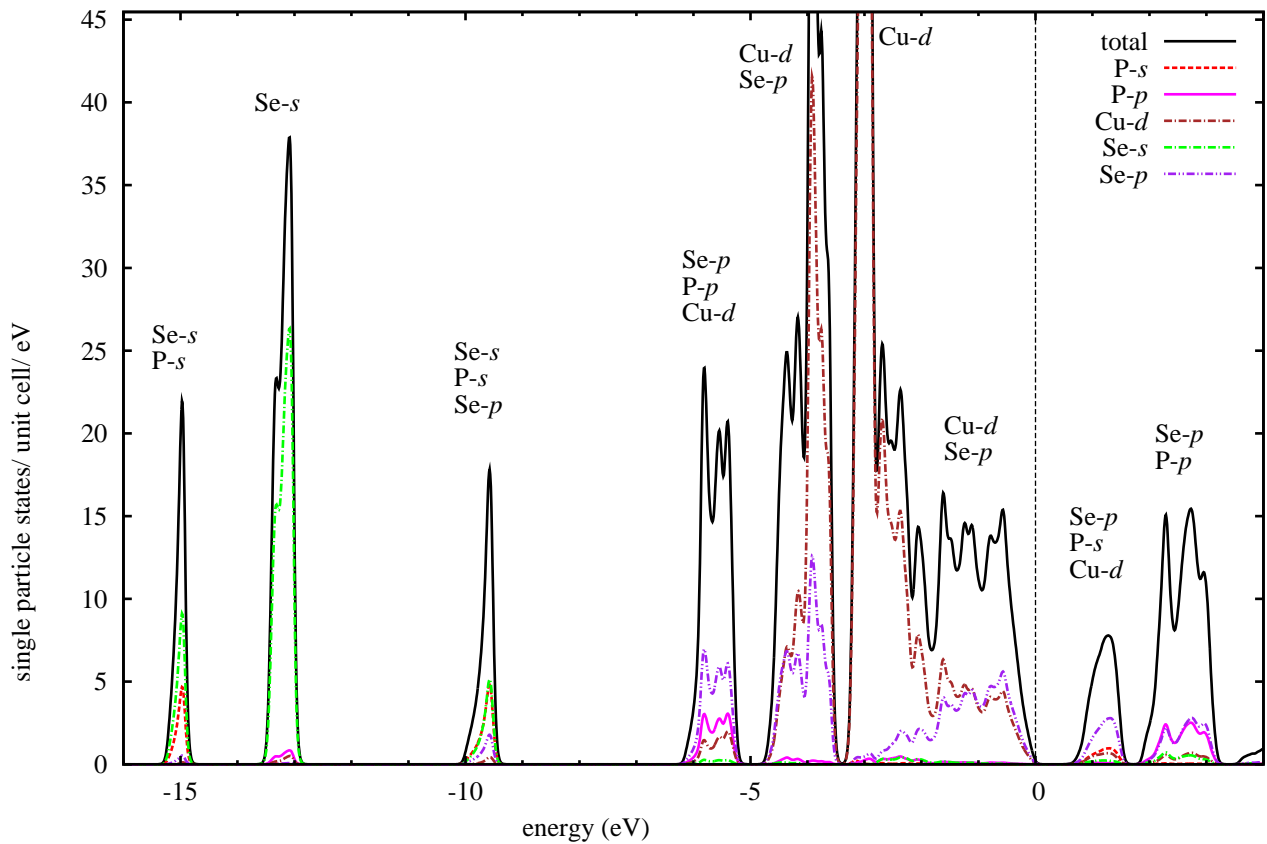}
\caption{\label{f:dos}(Color online) Partial density of states for the unit cell (Cu$_3$PSe$_4$)$_2$. The large Cu-$d$ peak rises to a maximum of twice the height of the plot. The vertical dashed line at 0 denotes the valence band maximum.}
\end{figure*}

\subsection{Defect Concentration}

Defect concentration is calculated in a two step process.
The first step self consistently solves for formation temperature concentrations $n_{D,q}^{(1)}$ of each defect type and charge state.
The second step self consistently solves for room temperature charge state concentrations $n_{D,q}^{(2)}$ while holding the defect type concentrations $n_D = \sum_q n_{D,q}$ fixed to their formation temperature values.
Nonequivalent sites of the defects V$_{\text{Cu}}$ and P$_{\text{Se}}$ have been treated as separate defects.
Multiplicities include equivalent site multiplicity and a spin degeneracy factor of two for charge states with half occupied orbitals.
For V$_{\text{Cu}}$ and P$_{\text{Se}}$, no other electronic degeneracies (or Jahn-Teller behavior) is present.
We use the full Fermi function rather than the Boltzmann approximation and calculated density of states distributions (with increased \hbox{$k$-point} density) rather than an effective density of state approximation based on effective masses.

\section{Results}

\subsection{Band Character}

The bonding character of Cu$_3$PSe$_4$~is evident in the GGA+$U$ calculation of the partial DOS, shown in Fig.~\ref{f:dos}.
The valence bands above $-7$ eV and the conduction bands below 3.3 eV have similarities to other multinary copper chalcogenides.
One such common property is that the Cu-$d$ states are split into non-bonding $e$ orbitals and $t_2$ orbitals which form filled bonding and filled antibonding bands because of their interaction with the chalcogenide $p$ orbitals\cite{zhang11com}.
The antibonding band forms the highest valence band.
Like CuInSe$_2$, CuGaSe$_2$, and Cu$_2$ZnSnSe$_4$\cite{zhang11com}, the conduction band has a character that is largely antibonding between Se-$p$ and Mt-$s$, where Mt represents the element acting as the high valence metal (e.g.~Sn in Cu$_2$ZnSnSe$_4$, P in Cu$_3$PSe$_4$).
The antibonding character is inferred from the presence of a spatial node between the Mt and Se atoms in the charge density of the lowest conduction band\cite{foster11ele}.
Unlike materials with a metallic Mt, Cu$_3$PSe$_4$~has no valence band that is the obvious bonding counterpart.
In fact, the P-$s$ orbitals have nominally been filled in the P-$s$/Se-$s$ bonding and antibonding bands, near $-15$ and $-10$ eV.
This $\sigma\sigma$ bonding does not occur when Mt is more metallic, because of the larger energy difference between the atomic Mt-$s$ level and the chalcogenide $s$ level.
Thus the appearance of a P-$s$/Se-$p^*$ antibond is somewhat surprising despite the fact that it follows the trend of other multinary copper chalcogenides.
The bonding counterpart of the \textit{second} conduction band, which has significant P-$p$/Se-$p^*$ character, is found in the valence band near $-5.7$ eV.

\subsection{Effective Mass and Dielectric Properties}

The calculated GGA+$U$ effective hole mass and dielectric tensor components are shown in Table \ref{t:masseps}.
The dielectric tensor is calculated using density functional perturbation theory\cite{Gonze97dyn}.
The effective mass tensor, calculated from the band structure, has much larger components in the $yz$ plane than along the $x$ axis.
Because the radius of a hydrogenic shallow defect state (also known as a perturbed host state\cite{lany08ass}) is inversely proportional to effective mass, this results in the shallow acceptor V$_{\text{Cu}}$ wavefunction being greatly elongated in the $x$ direction.
The conductivity effective hole mass is $m_{\text{cond}}^* \equiv 3/ \sum_i m_i^{-1} = 0.27$ $m_0$.
For comparison, the Si light and heavy hole effective masses are 0.16 and \hbox{0.49 $m_0$} respectively.

\begin{table}
\caption{\label{t:masseps}Principal axis tensor components and appropriate scalar averages for effective hole mass (units of electron mass $m_0$) and electronic and total dielectric constants, $\epsilon_{\infty}$ and $\epsilon_{0}$.}
\begin{tabular}{l|rrr|r}
\hline\\[-10pt]
& $x$ & $y$ & $z$ & scalar \\[2pt]
$m$ & 0.10 & 1.66 & 1.82 & $m^* = 0.67$,\quad$m_{\text{cond}}^* = 0.27$\\
$\epsilon$ (elect.) & 14.0 &  13.1 & 12.0 & $\epsilon_{\infty} = 13.0$ \\
$\epsilon$ (total) & 16.8 &  14.8 & 13.6 & $\epsilon_0 = 15.1$ \\ \hline
\end{tabular}
\end{table}

\subsection{Chemical Potential Domain}

\begin{figure}
\includegraphics{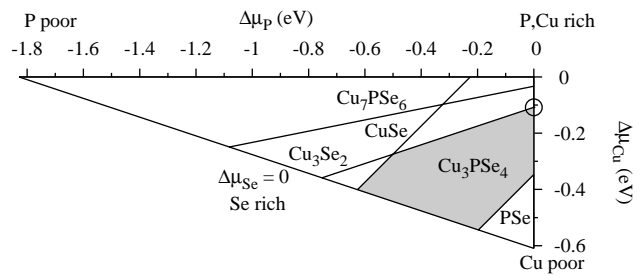}
\caption{\label{f:dmu}Chemical potential domain with stable region of Cu$_3$PSe$_4$~in gray. The chosen Cu-rich growth condition is indicated by a circle.}
\end{figure}

We analyze the allowed chemical potential domain for Cu$_3$PSe$_4$~synthesis by calculating $\Delta H$ for 22 compounds containing Cu, P, and Se.
Fig.~\ref{f:dmu} shows the results for several important compounds, revealing a
relatively large stable chemical potential domain.
To best match experimental carrier concentrations\cite{kokenyesi12tbd}, we perform the defect calculations for the conditions $\Delta \mu_{\text{P}}=0$ and $\Delta \mu_{\text{Cu}} = -0.11$ eV (circled in Fig.~\ref{f:dmu}).
Choosing $\Delta \mu_{\text{Cu}}$ to assume its maximum allowed value minimizes the calculated concentration of the shallow acceptor defect V$_{\text{Cu}}$.

We note that it has been observed\cite{Note4} that under certain conditions Cu$_3$PSe$_4$~can coexist with the ionic conductor Cu$_7$PSe$_6$, but this is not predicted by chemical potential domain analysis.
This discrepancy may be due to finite temperature effects; the low temperature $\alpha$ phase\cite{gaudin00str} of Cu$_7$PSe$_6$ was used in calculations, while at formation temperature the partially disordered $\gamma$ phase would be present.
We also note here that the error of total energy calculations involving phosphorus can be large; a statistical correction of 0.6 eV per P atom is given in Ref.~\onlinecite{lany08sem} due to artefactual energy differences between phosphorus in reductive and neutral (elemental) environments.
This error is expected to impart uncertainty both to the calculated heat of formation of Cu$_3$PSe$_4$, which affects defect energies through its effect on $\Delta \mu_{\text{Cu}}$, and to the defect supercell energies themselves, particularly for the high concentration P$_{\text{Se}}$ defect.
In the latter case, the additional P atom is reduced by the neighboring Cu ions, in strong contrast to the host P atoms, which are oxidized by the Se neighbors.
While phosphorus raises concern, the GGA+$U$ statistical corrections\cite{lany08sem} associated with Cu and Se atoms are less than 0.05 eV, and our calculated heat of formation of Cu$_3$Se$_2$ is within 0.05 eV of experiment\cite{lany08sem}.

\subsection{Defect Analysis}

The defect analysis is performed initially using a $2^3$ supercell (128 atoms).
We use all finite size corrections described above ($E_{\text{BF}}$, $E_{\text{PA}}$, $E_{\text{MP}}$).
For the GGA+$U$ calculation, no correction is made to the valence band, while the conduction band correction \hbox{$\Delta E_C = 0.88$ eV} is obtained from the difference of the experimental bandgap (1.4 eV) and the calculated bandgap (0.52 eV).
A shallow donor correction term is applied to the energies of incompletely ionized shallow donor defects.
However, none of the intrinsic point defects are clearly shallow donors, and thus this correction is applied only for the extrinsic donors considered: Ca, Cd, and Zn on a Cu site, and Cl on a Se site.
(Here the correction is $+(1-q) \Delta E_C$, since $z_e=1-q$ with  $q=0,1$.)

\begin{figure}
\includegraphics{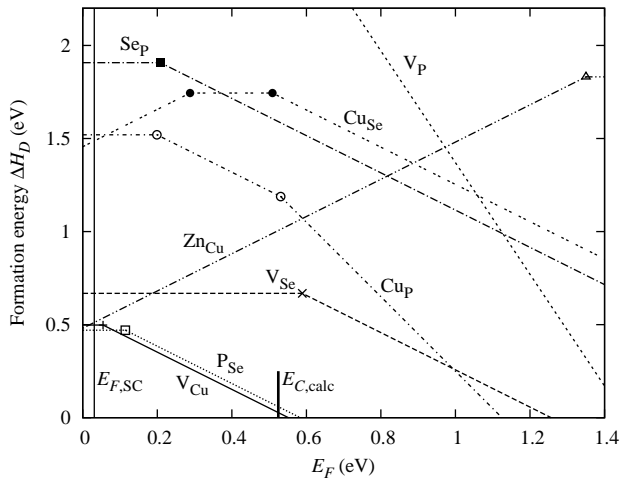}
\caption{\label{f:defects}Defect formation energies and transition energies. Where nonequivalent sites are calculated, the lowest energies for each charge state are shown.
The self-consistent room temperature Fermi energy $E_{F,\text{SC}}$, assuming a formation temperature of 500$^\circ$C, is shown by the vertical line at 0.031 eV. The (0/+) transition energy for shallow donor Zn$_{\text{Cu}}$ has been raised to follow the conduction band correction.}
\end{figure}

Formation energies and transition energies for the lower energy intrinsic defects and the lowest energy extrinsic defect are shown in Fig.~\ref{f:defects}.
 The acceptors V$_{\text{Cu}}$ and  P$_{\text{Se}}$  both pin the Fermi energy below mid-gap, preventing Cu$_3$PSe$_4$~from being $n$-doped near thermal equilibrium.
The formation energy of the neutral defect V$_{\text{Cu}}^0$ is calculated to be 0.50 eV, with a (-/0) transition energy of 0.05 to 0.06 eV, depending on the Cu site.
The formation energy of P$_{\text{Se}}^0$ varies with site from 0.47 eV to 0.50 eV, with the (-/0) transition energies varying from 0.08 eV to 0.17 eV.

The net finite size corrections for the two important intrinsic defects V$_{\text{Cu}}$ and P$_{\text{Se}}$ lie between 0 and \hbox{0.11 eV}, with the potential alignment correction ranging from  0 to \hbox{$0.06$ eV}.
For these defects, only the charge neutral defect states have non-zero $\Delta E_{\text{BF}}$ values (0 to \hbox{$-0.04$ eV}).
A band filling correction analogous to Eq.~(\ref{e:bf}) is used for the extrinsic donor defect Zn$_{\text{Cu}}$, yielding \hbox{$\Delta E_{\text{BF}} = -0.08$ eV} for the neutral defect state.

We use the formation temperature of 500$^{\circ}$C (approximately the temperature used in recent pellet and single crystal experiments\cite{kokenyesi12tbd}) to calculate the concentrations of each defect type.
The resulting defect concentrations (irrespective of charge state) are \hbox{$4.1 \times 10^{19}$ cm$^{-3}$} for V$_{\text{Cu}}$ and \hbox{$4.6 \times 10^{19}$ cm$^{-3}$} for P$_{\text{Se}}$.
The second step of the concentration calculation yields a room temperature (300 K) self-consistent Fermi level of \hbox{0.031 eV} above the VBM and a hole concentration of \hbox{$p = 8 \times 10^{18}$ cm$^{-3}$}.
V$_{\text{Cu}}$ is electronically the most important defect type, since the contribution of V$_{\text{Cu}}^{-}$ to the hole density is over five times that of P$_{\text{Se}}^{-}$.

If zinc is present during 
synthesis, the maximum Zn$_{\text{Cu}}^{+}$ concentration is approximately $5\times10^{18}$ cm$^{-3}$, and the net room temperature hole density is lowered slightly to $p= 6\times10^{18}$ cm$^{-3}$.
The other potential donor dopants considered have greater formation energies and can be neglected for all growth conditions.

We have recalculated the charged configurationsof the weakly localized V$_{\text{Cu}}$ defect using a $4^3$ (1024 atom) supercell.
Even for this supercell size, the defect wavefunction is not localized within the supercell in the $x$ direction (the low effective mass direction).

For the neutral defect 
the calculated (-/0) charge transition energy is \hbox{0.04 eV} and \hbox{$\Delta H = 0.53$ eV} .
We note that the hydrogen-like approximation using the conductivity effective mass yields a comparable binding energy of \hbox{0.02 eV}.
Assigning the large supercell data to all Cu sites, in combination with the previous P$_{\text{Se}}$ data, yields an insignificantly modified hole density \hbox{$p= 9\times10^{18}$ cm$^{-3}$}.

The P$_{\text{Se}}$ defect state is substantially localized within the smaller $2^3$ supercell.
The defect state has P-$p$ character on the defect (Se) site and Cu-$d$ character on the nearest neighbors, similar to a localized version of the host valence band, which has Se-$p$ and Cu-$d$ character.
The degree of localization allows us to apply the defect image charge correction of Refs.~\onlinecite{freysoldt10ele,*freysoldt09ful} using the neutral defect potential as the reference potential (see Appendix \ref{as:Freysoldt}).
The resulting correction (0.08 eV) agrees well with the corresponding correction ($\Delta E_{\text{MP}} + \Delta E_{\text{PA}} = 0.07$ eV) according to Refs.~\onlinecite{lany08ass,lany09acc}.

\begin{table}
\caption{\label{t:GGAUandGGAandExp}Calculated site-averaged formation energies for V$_{\text{Cu}}$ defects and predicted versus experimental hole concentrations.
The error of the GGA+$U$ method is seen to be smaller than the more standard\cite{lany08ass} method of using GGA including VBM corrections.}
\begin{tabular}{l|l|l|l}
\hline\\[-8pt]
 & $\Delta H$(V$_{\text{Cu}}^0$) [eV] & $\Delta H$(V$_{\text{Cu}}^{-}$) & $p$ [cm$^{-3}$] \\[2pt]
GGA + VBM corr. & 0.34 & 0.40 & $6\times10^{19}$ \\
GGA+$U$ & 0.50 & 0.56 & $8\times10^{18}$ \\
Hall measurement\cite{kokenyesi12tbd} & {---} & {---} & $6\times10^{17}$ \\
\hline
\end{tabular}
\end{table}

\subsection{Discussion and Further Investigation}

The defect analysis performed here agrees qualitatively with recent experimental results.
Our calculated Cu:P ratio of 2.97 is consistent with the value $2.92 \pm 0.06$ measured for single crystals\cite{kokenyesi12tbd}.
We predict a large hole concentration of $p = 8 \times 10^{18}$ cm$^{-3}$, about one order of magnitude larger than the value $6 \times 10^{17}$ cm$^{-3}$ obtained by Hall and Seebeck measurements on pressed, sintered pellets\cite{kokenyesi12tbd}.

We compare the GGA+$U$ defect calculations described above with standard GGA defect calculations followed by application of a GGA+$U$ correction\cite{persson05ntd,lany08ass} ($-0.34$ eV) to the VBM.
The GGA defect calculations include all types of corrections applied to the GGA+$U$ calculations and include a GGA determination of the maximum allowed copper chemical potential $\Delta \mu_{\text{Cu}}$ ($-0.06$ eV).
As shown in Table \ref{t:GGAUandGGAandExp}, the more standard ``GGA + VBM correction''  procedure changes the formation energies of V$_{\text{Cu}}^0$ and V$_{\text{Cu}}^-$ (evaluated at maximum $\Delta \mu_{\text{Cu}}$ and minimum $E_{F}$) by about $-0.16$ eV, causing a significantly larger overestimation of $p$ relative to reported experimental values. 
This comparison shows that GGA+$U$ performs better than GGA not only in bulk total energy calculations\cite{lany08sem,stevanovic12cor}, but also in defect calculations\cite{{scanlon09mod}}.

It is instructive to consider further the implications of the available experimental results\cite{kokenyesi12tbd}.
We examine possible changes in defect formation enthalpies which would
bring the calculated hole concentration $p$ closer to the value measured for polycrystalline pellets\cite{kokenyesi12tbd}.
If one assumes that the calculated transition energy of V$_{\text{Cu}}$ is not underestimated, the experiments of Ref.~\onlinecite{kokenyesi12tbd} indicate that the formation energy of V$_{\text{Cu}}$ must increase, while the transition energy of P$_{\text{Se}}$ increases and the formation energy of P$_{\text{Se}}$ decreases.
The adjustment to the V$_{\text{Cu}}$ energy must be significant to recover the measured $p$.
For example, increasing the formation energy of V$_{\text{Cu}}$ defects by 0.35 eV while applying changes of $-0.05$ and 0.05 eV to the neutral and charged P$_{\text{Se}}$ defects respectively yields $p = 7\times10^{17}$ cm$^{-3}$ and a Cu:P ratio of 2.96.
Such large changes to the V$_{\text{Cu}}^q$ formation energies 
cannot readily be explained by systematic
calculational errors associated primarily with phosphorus.

An alternative possibility is that the GGA+$U$ calculated VBM is too high by a modest amount, and that the apparent shallow character of the V$_{\text{Cu}}$ defect is an artifact of this band misplacement.
For example, applying a valence band correction $\Delta E_V = - 0.1$ eV and choosing not to apply the shallow acceptor corrections to the neutral defects (that is, using a strictly ``band edge only'' approach) yields the much lower hole concentration $p=1.1 \times 10^{18}$ cm$^{-3}$ with an only slightly increased Cu:P ratio (2.976).

\begin{figure}
\includegraphics{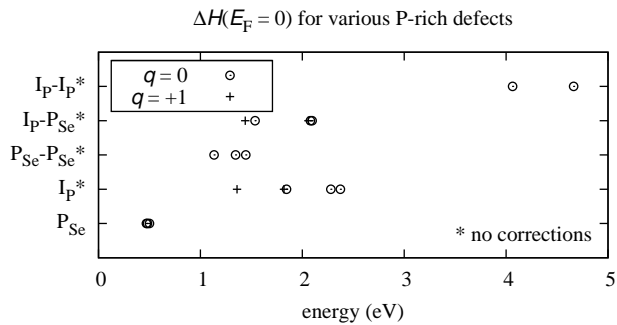}
\caption{\label{f:Pdef}Heat of formation (GGA+$U$) of interstitial and complex defects containing extra P atoms.
Energy corrections $\Delta E_{\text{corr}}$ are neglected except for the P$_{\text{Se}}$ defect.}
\end{figure}

The experimental data suggests an increase in neutral P concentration, and possibly the presence of a low energy \textit{donor} defect involving extra P atoms. 
Such a donor defect could lower 
the hole concentration by compensating the V$_{\text{Cu}}$ acceptors and thus avoid the need to raise $\Delta H$(V$_{\text{Cu}}$).
We therefore have examined, at lower accuracy and without finite size corrections, a number of neutral and positively charged P-rich defects, including interstitials and complexes in various configurations.
The results are shown in Fig.~(\ref{f:Pdef}).
The energies suggest that there are no significant sources of extra phosphorus besides P$_{\text{Se}}$.

\section{Conclusion}

In conclusion, we have performed a set of GGA+$U$ defect calculations on Cu$_3$PSe$_4$, a $p$-type semiconductor with a direct bandgap of 1.4 eV.
We compare our methods against standard GGA, larger supercells, and alternative correction methods.
We predict that the V$_{\text{Cu}}$ defect is mostly responsible for the large, experimentally observed intrinsic hole concentration $p$, with some contribution from P$_{\text{Se}}$.
Both of these defects pin the Fermi level below mid-gap, so that $n$-doping is  prohibited near thermal equilibrium.
Both defects also contribute to the observed non-stoichiometric Cu:P ratio.
Our calculation overestimates 
the hole concentration $p$ by about one order of magnitude.
Overall, the GGA+$U$ method is shown to be more accurate than standard GGA calculations with valence band corrections.
Doping with Zn is calculated to have a small but noticeable effect on $p$.
Because of the apparent uncertainty in the calculations however, this analysis does not rule out the possibility that Zn doping could significantly reduce $p$.

\begin{acknowledgments}
We thank Dr.~Robert Kokenyesi from Oregon State University (OSU) for helpful discussions on both experimental and theoretical topics.
We thank Dr.~Stephan Lany from the National Renewable Energy Laboratory for very helpful discussions and for providing scripts with which some of the potential alignment and band filling corrections are calculated.
This work has been supported by the National Science Foundation of the USA under Grant SOLAR DMS-1035513.
\end{acknowledgments}

\appendix
\section{\label{as:GGAVBMcorr}The GGA + VBM correction calculations}
Except for the final VBM correction, the chemical potential domain analysis and defect analysis for the GGA calculations are performed with $U=0$.
Similar to the set of GGA+$U$ calculations, bulk relaxations are performed for the unit cells of elements Cu, P, Se, and relevant compounds such as Cu$_3$Se$_2$ and PSe, in order to obtain the chemical potential domain and determine the maximum $\Delta \mu_{\text{Cu}}$.
Relaxed lattice parameters for Cu$_3$PSe$_4$ are also recalculated with $U=0$ in order to create the $2^3$ supercell for the GGA defect calculations.
Defect calculations are performed with the same types of corrections as are used for the wholly GGA+$U$ method.

The value of $\Delta E_V$ is determined in the following manner.
A static (ion-fixed) GGA+$U$ calculation of bulk Cu$_3$PSe$_4$~is performed using the GGA-relaxed unit cell.
The energy of the resulting VBM relative to the mean energy of the Se-$s$ peak (used as a reference) is taken to be the relative VBM of the GGA + VBM correction method.
This energy, minus the corresponding relative VBM of the plain GGA unit cell calculation, gives the valence band correction $\Delta E_V$:
\begin{align}
\Delta E_V =& [E_V^{\text{GGA + VBM corr.}} - E_{\text{Se-}s}^{\text{GGA + VBM corr.}}] \notag \\
&- [E_V^{\text{GGA}} - E_{\text{Se-}s}^{\text{GGA}}]
\end{align}
We find that an alternate reference, the average electrostatic potential of spheres centered on the Se atoms, results in negligible differences.

\section{\label{as:Freysoldt}Alternative ``Model charge'' electrostatic corrections}
We performed the alternative electrostatic $+$ potential alignment correction as  described in Refs.~\onlinecite{freysoldt10ele,*freysoldt09ful}, with the exception that the electrostatic potential from the neutral defect calculation was used as a reference potential, instead of the potential of the host supercell.
This was necessary in order to locate the potential asymptote away from the defect.
When the host potential was used as a reference, the motion of the ions upon relaxation caused extreme oscillations in the potential difference $V_{D,q} - V_H$.
The electrostatic potential difference $V_{D,q} - V_{D,0}$ on the other hand, involved much less radical oscillations due to the relative ionic motion, and allowed the asymptote to be located.
(This issue was avoided in Refs.~\onlinecite{freysoldt10ele,*freysoldt09ful} by not allowing ionic relaxation.)

Operationally, we constructed a model, periodic, spherical Gaussian + exponential charge distribution on the same real space lattice that was assumed by the 
DFT/PAW charge and potential distributions.
Periodicity was realized through the Fourier representation of the functions.
The Gaussian width parameter $\beta$ was set to 2 bohr, as done in Refs.~\onlinecite{freysoldt10ele,*freysoldt09ful}.
The ratio $x$ of exponential to Gaussian character was parametrized by a periodic variable $t$.
The parameter $t$, the exponential width parameter $\gamma$ and the center of the charge distribution $(x_0, y_0, z_0)$ were fit to the defect charge state using a constraint on $\min(\gamma)$.
The program \texttt{sxdefectalign} by Christoph Freysoldt was then used to obtain the final corrections\cite{freysoldt10ele,*freysoldt09ful}.


%

\end{document}